# Equivalent Strain in Simple Shear Deformations

Yan Beygelzimer

Donetsk Institute of Physics and Engineering The National Academy of Sciences of Ukraine

**Abstract**

We show that simple shear and pure shear form two groups of transformations with different properties. The equivalent strain is viewed as an external control parameter of the deformation process at low homologous temperatures. The von Mises strain satisfies group-theoretic properties of both groups, supporting its use for measuring the equivalent strain. The Hencky strain, on the other hand, does not satisfy the simple shear group properties, implying that it is not appropriate for measuring the equivalent strain in simple shear.

The paper also proposes a hypothesis explaining the absence of metal hardening in large simple shear deformations. This hypothesis explains why excluding rotations as prescribed by the finite strain theory is not valid.

## 1. Introduction

Fueled by advances in severe plastic deformation [1], there has been much recent discussion about computing the equivalent strain $e$ in simple shear. Since the increment of equivalent strain can be defined via the increment of plastic work, the authors of [2 - 5] argue for using the von Mises strain

$$e_M = \frac{\gamma}{\sqrt{3}}, \qquad (1)$$

where $\gamma$ is the shear strain. Onaka [6, 7], on the other hand, argues that one must use the Hencky relation

$$e_H = \frac{2}{\sqrt{3}} \ln\left[\left(1 + \frac{\gamma^2}{4}\right)^{1/2} + \frac{\gamma}{2}\right] \qquad (2)$$

The justification given by Onaka is the need to exclude rotations from the deformation-gradient tensor in large simple shear deformations.

There is another line of reasoning that led some authors (see, e.g., [8]) to formula (2). Starting with Ludwik [9], it was widely believed that the same value of $e$ guarantees essentially the same level of hardening, regardless of the deformation mode. Seen in this light, the estimate (1) is unreasonably high if one compares the hardening achieved by simple shear and that achieved by other deformation modes (e.g., elongation).

In [10], the equivalent strain is defined as a measure of hardening. This definition gave an expression for the equivalent strain in simple shear predicting that $e$ saturates at large values of $\gamma$. This saturation is in agreement with an empirically observed fact that, in large deformations caused by high pressure torsion, materials reach a stationary structure without hardening [11 - 13]. When viewed as a measure of hardening, the equivalent strain becomes a physical (instead of a geometric) parameter that depends on the material's properties.

Here we take a synergetics approach [14], viewing the equivalent strain as an external control parameter of the deformation process. Using group-theoretic properties of geometric transformations [15] and the principle of additivity, we show that the equivalent strain of simple shear must be linear in $\gamma$. This approach gives evidence in favor of using the von Mises strain for computing the equivalent strain in simple shear.

The paper also proposes a hypothesis explaining the absence of metal hardening in large deformations caused by simple shear. This hypothesis explains why excluding rotations as prescribed by the finite strain theory is not valid.

## 2. The von Mises strain and group theoretic properties of simple and pure shear

We view the equivalent strain as an external scalar control parameter determining the deformation process at low homologous temperatures. We will show that, for simple shear, such a control parameter must be a linear function in shear strain $\gamma$. For this, consider simple shear as a geometric transformation defined by relation

$$\mathbf{x} = \mathbf{S}\mathbf{X} , \tag{3}$$

where $\mathbf{X}$ and $\mathbf{x}$ are the initial and the final coordinates of a point, respectively; and

$$\mathbf{S} = \begin{bmatrix} 1 & \gamma & 0 \\ 0 & 1 & 0 \\ 0 & 0 & 1 \end{bmatrix} \tag{4}$$

is the transformation operator.

It is easily verified that the set of all simple shear transformations along a fixed shear direction satisfies the four axioms of group theory [15]. Indeed, consider two consecutive transformations (4) with shear strains $\gamma_1$ and $\gamma_2$ respectively:

$$\mathbf{S}_1 = \begin{bmatrix} 1 & \gamma_1 & 0 \\ 0 & 1 & 0 \\ 0 & 0 & 1 \end{bmatrix} = \mathbf{S}(\gamma_1) \text{ and } \mathbf{S}_2 = \begin{bmatrix} 1 & \gamma_2 & 0 \\ 0 & 1 & 0 \\ 0 & 0 & 1 \end{bmatrix} = \mathbf{S}(\gamma_2) \tag{5}$$

It is clear that

$$\mathbf{S}_2\mathbf{S}_1 = \begin{bmatrix} 1 & \gamma_2 & 0 \\ 0 & 1 & 0 \\ 0 & 0 & 1 \end{bmatrix}\begin{bmatrix} 1 & \gamma_1 & 0 \\ 0 & 1 & 0 \\ 0 & 0 & 1 \end{bmatrix} = \begin{bmatrix} 1 & \gamma_1+\gamma_2 & 0 \\ 0 & 1 & 0 \\ 0 & 0 & 1 \end{bmatrix} = \mathbf{S}(\gamma_1+\gamma_2) \tag{6}$$

Thus a composition of two simple shear transformations along the same direction is a simple shear transformation along that direction, and the set defined by (4) satisfies the axiom of closure. Similarly, it is easy to see that (4) satisfies the associativity axiom, $(\mathbf{S}_3(\mathbf{S}_2\mathbf{S}_1) = (\mathbf{S}_3\mathbf{S}_2)\mathbf{S}_1)$, contains the identity element $\mathbf{I} = \mathbf{S}(0)$, and that for every transformation $\mathbf{S}(\gamma)$ it contains the inverse $\mathbf{S}^{-1}(\gamma) = \mathbf{S}(-\gamma)$ such that $\mathbf{S}^{-1}(\gamma)\mathbf{S}(\gamma) = \mathbf{I}$.

We will associate each transformation $\mathbf{S}(\gamma)$ with the value of its characteristic control parameter, the equivalent strain $e_s = e_s(\gamma)$, where the subscript "s" stands for "simple shear". Consider two consecutive simple shear transformations $\mathbf{S}(\gamma_1)$ and $\mathbf{S}(\gamma_2)$ with corresponding equivalent strains

$$e_{s1} = e_s(\gamma_1) \text{ и } e_{s2} = e_s(\gamma_2). \tag{7}$$

It follows from (6) that the total shear is characterized by the equivalent strain

$$e_{s\Sigma} = e_s(\gamma_1 + \gamma_2). \tag{8}$$

On the other hand, the equivalent strain is additive [16]; this means that

$$e_{s\Sigma} = e_{s1} + e_{s2} \tag{9}$$

Plugging in expressions (7) and (8), we get

$$e_s(\gamma_1) + e_s(\gamma_2) = e_s(\gamma_1 + \gamma_2) \tag{10}$$

A solution to this functional equation is a linear function (see, e.g., [17])

$$e_s = C\gamma, \tag{11}$$

for some constant $C$.

To match the von Mises values at low equivalent strain [16], $e_s(\gamma) = e_M = \dfrac{\gamma}{\sqrt{3}}$ yielding $C = \dfrac{1}{\sqrt{3}}$.

The derivation of relation (11) does not require $\gamma$ to be small, and so $e_s = \dfrac{1}{\sqrt{3}}\gamma$ is valid for any value of $\gamma$, agreeing with equation (1).

This linear dependence of the equivalent strain from $\gamma$ follows from group-theoretic properties of simple shear and the additivity of the equivalent strain. A similar argument gives a logarithmic dependence in the case of constant-volume pure shear. Indeed, the transformation operator in this case has the form

$$\mathbf{P}(\lambda) = \begin{bmatrix} \lambda & 0 & 0 \\ 0 & \lambda^{-1} & 0 \\ 0 & 0 & 1 \end{bmatrix}, \tag{12}$$

where $\lambda$ is the relative elongation.

One can easily verify that the set of pure shear transformations forms a group whose elements satisfy the relation

$$\mathbf{P}(\lambda_2)\mathbf{P}(\lambda_1) = \mathbf{P}(\lambda_2\lambda_1) \tag{13}$$

Unlike the analogous relation for simple shear (6), it is nonlinear in the sense that the parameter of the total transformation is the product instead of the sum of the two components. It is this difference in group-theoretic properties of simple and pure shear that leads to a different dependence of the equivalent strain from $\gamma$ in these processes.

It follows from relation (13) and the additivity property that the equivalent strain $e_p$ (where "p" refers to "pure shear") satisfies equation

$$e_p(\lambda_1) + e_p(\lambda_2) = e_p(\lambda_1 \lambda_2) \tag{14}$$

instead of equation (10) as in the case of simple shear.

A solution of this functional equation is the logarithm function (see, e.g., [17]),

$$e_p(\lambda) = K \ln \lambda, \tag{15}$$

where $K$ is a constant.

This dependence of the equivalent strain on the relative elongation corresponds to an expression for the von Mises strain, which in the case of pure shear is determined by the following formula [16]:

$$e_M(\lambda) = \frac{2}{\sqrt{3}} \ln \lambda \tag{16}$$

In this case, the von Mises strain again follows from group-theoretic properties and the additivity principle.

## 3. Discussion

The previous section argued for using the von Mises strain in both pure and simple shear, without relying on any model of the material and without making any assumptions on the shear strain. Thus the von Mises strain can be used as a control parameter not only in the case of isotropic bodies where $de_M$ defines the incremental work, but also in other settings—for example, in anisotropic bodies [16] or in the case of strain gradient plasticity [18].

Two questions arise in light of Section 2 and the discussion [2-8]:

1. Why does strain hardening in simple shear differ substantially from that in pure shear, for the same value of the von Mises strain, when the von Mises strain becomes sufficiently large?

2. Why does excluding rotations, as dictated by the finite strain theory, lead to a wrong result?

To answer these questions, we will first try to answer the contrapositive of question 1: A polycrystal specimen represents a system with multitudinous degrees of freedom. So why does a single scalar control parameter determine hardening and the average grain size (at relatively small equivalent strains and low homologous temperatures)? This is surprising, like other examples of simple behavior in complex systems, such as turbulent flows in liquids.

Kolmogorov [19] hypothesized the structural self-similarity of turbulent flows and its defining scaling laws. Barenblatt [20] gives other examples of simple behavior of complex systems related to their self-similarity. We believe that the answer to the contrapositive lies precisely in the self-similarity of metal structure evolution during deformation. Such self-similarity has been observed experimentally for sufficiently large strains (see, for example, [21]). It is also the self-similarity that is responsible for the power low

$$\sigma = f(e) \tag{17}$$

of the equivalent stress $\sigma$ from the equivalent strain $e$, repeatedly confirmed in experiments

(see, e.g., [16]). In other words, the stress-strain curve is an expression of scaling behavior, which, according to [20], is common to all self-similar processes.

We will show that $f(e)$ is an power low during the self-similar stage of microstructure evolution. Consider three consecutive states of the system, specified by $(e_1;\sigma_1)$, $(e_2;\sigma_2)$, and $(e_3;\sigma_3)$. Since the evolution of metal structure is self-similar, the estimate (17) is invariant with respect to the choice of unit for measuring $e$. Let us first choose $e_1$ as the unit measure. Then $f(e)$ must have the following form:

$$f(e) = \sigma_1 \varphi\left(\frac{e}{e_1}\right), \tag{18}$$

where $\varphi(1) = 1$.

For states $(e_2;\sigma_2)$ and $(e_3;\sigma_3)$, according to (18), we get

$$\frac{\sigma_2}{\sigma_1} = \varphi\left(\frac{e_2}{e_1}\right), \tag{19}$$

$$\frac{\sigma_3}{\sigma_1} = \varphi\left(\frac{e_3}{e_1}\right), \tag{20}$$

If we choose $e_2$ as our unit measure, a similar argument will lead to the following relation:

$$\frac{\sigma_3}{\sigma_2} = \varphi\left(\frac{e_3}{e_2}\right). \tag{21}$$

Multiplying (19) and (21) and comparing with (20), we get

$$\varphi\left(\frac{e_3}{e_1}\right) = \varphi\left(\frac{e_3}{e_2}\right) \cdot \varphi\left(\frac{e_2}{e_1}\right) \tag{22}$$

This implies that $\varphi(x)$ satisfies the following equation

$$\varphi(x_1 \cdot x_2) = \varphi(x_1) \cdot \varphi(x_2), \tag{23}$$

A solution to such an equation is an power low (see, for example, [17]):

$$\varphi(x) = x^n, \tag{24}$$

where $n$ is a parameter. Thus

$$\sigma = A e^n, \tag{25}$$

where $A$ is a parameter.

Grain refinement can be viewed as recursive grain subdivision [22]. Kolmogorov [23] investigated a fairly general model of recursive particle subdivision, showing that self-similar structures emerge at a certain stage of this process—if the subdivision mechanism is constant and scale-invariant. Other research [21, 24 - 26] showed the universality and scale-invariance of grain refinement during plastic deformation. This leads to the conclusion that as long as the deformation mechanism is unchanged, metal structure will evolve in a self-similar manner, resulting in a universal power low stress-strain curve. Thus a deviation from this curve must be related to a change in the deformation mechanism.

Papers [10 ,27] make a case for a hypothesis that a certain percolation mechanism sets in during simple shear. This mechanism explains the lack of deformation hardening (question 1 above) and the formation of a stationary microstructure observed experimentally [11-13], as well as a number of other phenomena in large simple shear deformations [28].

The answer to question 2, we believe, is related to the same percolation mechanism. According to [27], this mechanism has a relaxation nature, periodically relieving internal stress via small rotations of grain clusters. This process allows the material to repair its structure. This way, the deformed material guarantees a stationary character of transformation (4) expressed using property (6). Large simple shear deformations are realized as a sum of small independent steps, explaining the special character of rotations in this setting.

## 4. Conclusion

Simple and pure shear deformations form two groups of geometric transformations with different properties. The von Mises strain satisfies group-theoretic properties of both simple and pure shear. The Hencky strain, on the other hand, does not satisfy the properties of the simple shear group. This brings evidence that the von Mises strain is the correct measure of the equivalent strain in simple shear.

Since the von Mises strain is justified using group-theoretic arguments, without replying on any model of the material, this points to the applicability of the von Mises strain as a control parameter not only in isotropic bodies but in other settings as well.